\documentclass[12pt,final]{elsart}

  \usepackage{latexsym,bm,amsmath,amssymb,amsfonts}
  \usepackage{epsfig,graphics,graphicx}
  \usepackage{slashed}
\usepackage{latexsym,bm,amsmath,amssymb,amsfonts}

\newcommand{\x}{{\bm x}}
\newcommand{\y}{{\bm y}}
\newcommand{\z}{{\bm z}}
\newcommand{\n}{{\bm n}}
\newcommand{\U}{\tilde{U}}

\newcommand{\tr}{{\rm tr}  }

\newcommand{\beq}{\begin{eqnarray}}
\newcommand{\eeq}{\end{eqnarray}}

\begin{document}

\begin{flushright}
YITP13-30\\
TTP13-012\\
SFB/CPP-13-25\\
\vspace{5mm}
\end{flushright}

\begin{frontmatter}

\parbox[]{16.0cm}{ \begin{center}
\title{Resummation of non-global logarithms at finite $N_c$}

\author{Yoshitaka Hatta$^{\rm a}$ and Takahiro Ueda$^{\rm b}$ }


\address{$^{\rm a}$ Yukawa Institute for Theoretical Physics, Kyoto University, Kyoto 606-8502, Japan}
\address{$^{\rm b}$ Institut f\"{u}r Theoretische Teilchenphysik, Karlsruhe Institute of Technology (KIT), \\ D-76128 Karlsruhe, Germany }
\end{center}


\begin{abstract}

In the context of inter-jet energy flow, we present the first quantitative result of the resummation of non-global logarithms at finite $N_c$. This is achieved by refining Weigert's approach in which the problem is reduced to the simulation of  associated Langevin dynamics in the space of Wilson lines. We find that, in $e^+e^-$ annihilation, the exact result is rather close to the result previously obtained in the large--$N_c$  mean field approximation. However, we observe  enormous event--by--event fluctuations in the Langevin process which may have significant consequences in hadron collisions.
\end{abstract}
}

\end{frontmatter}

\section{Introduction}

In certain search channels of the Higgs boson and new particles at the Large Hadron Collider (LHC), it is often desirable to be able to control QCD radiation from tagged jets in order to suppress large backgrounds. A prime example is the Higgs boson production in association with di-jets. The two competing production mechanisms, the gluon fusion and the vector boson fusion processes, have  different patterns of soft gluon radiation due to the difference in their color structure.  Discriminating these processes quantitatively using some measure of radiation is therefore a useful strategy to determine the Higgs couplings \cite{Cox:2010ug,Englert:2012ct}.

 A related class of observables which are particularly sensitive to soft radiation is the cross section with a veto on unwanted jets in the full or partial region of the phase space. This generally requires the resummation of logarithms in $p_T^{\rm veto}$, the threshold transverse momentum of vetoed jets. Steady progress in this direction has been made for \emph{global} observables which involve all the particles and jets in the final state including those close to the beam axis.  The state--of--the--art is that one can resum the leading logarithms (LL) $(\alpha_s \ln^2 p_T^{\rm veto})^n$, the next--to--leading logarithms (NLL) $(\alpha_s \ln p_T^{\rm veto})^n$ and even the next--to--next--to--leading logarithms (NNLL) $(\alpha_s^2\ln p_T^{\rm veto})^n$ \cite{Berger:2010xi,Becher:2012qa,Banfi:2012jm}.

   However, in contrast to such progress, there exists a severe limitation in our   ability to resum \emph{non-global} logarithms \cite{Dasgupta:2001sh,Dasgupta:2002bw} which arise when measurements are restricted to a part of the phase space excluding the beam and jet regions. In this case,
   double--logarithms $(\alpha_s \ln^2 p_T^{\rm veto})^n$ are absent due to the lack of the collinear singularity. The leading contribution is then comprised of single--logarithms $(\alpha_s \ln p_T^{\rm veto})^n$ which originate from the soft singularity. The problem is that these logarithms do not exponentiate, and because of this difficulty their resummation has been hitherto  done only in the large--$N_c$ limit \cite{Dasgupta:2001sh,Dasgupta:2002bw,Banfi:2002hw,Marchesini:2003nh}. In other words, even the leading logarithms cannot be fully satisfactorily resummed. This could be a potentially serious drawback in actual experiments considering the  fact that, strictly speaking, any vetoed cross section at hadron colliders is inevitably non-global due to the finite acceptance of detectors.

 In fact, there is a single work by Weigert \cite{Weigert:2003mm} which did discuss the resummation of non-global logarithms at finite $N_c$ in a simpler setup of $e^+e^-$ annihilation where complicacies from the initial state radiation do not arise. His approach is based on an analogy with another, seemingly unrelated resummation  in QCD, namely, that of the small--$x$ (or `BFKL') logarithms in Regge scattering. In this context, a very similar issue arises as to how one can generalize the equation which resums  small--$x$ logarithm in the large--$N_c$ limit to one at finite $N_c$.  It turns out that these equations bear a striking resemblance to the  equation  which resums non-global logarithms in the large--$N_c$ limit. Since technologies to solve the former problem are well--developed, they may be suitably adapted to address the latter problem as well.
Somewhat surprisingly, however, Weigert's approach has not been pursued for a decade. Part of the reason of this may perhaps be  that, as we shall point out, there is actually a flaw in his formulation which deters a straightforward numerical implementation. In this paper we overcome this difficulty and  present the first quantitative results of the resummation of non-global logarithms at finite $N_c$.

 In Section 2, we quickly review the nonlinear evolution equations which resum the small--$x$ logarithms in high energy QCD. The subject may seem utterly unfamiliar to the readers whose primary interest is   jet physics. However, the similarity (or even equivalence) to the resummation of non-global logarithms will soon become apparent in Section 3 where we introduce the relevant evolution equations. Using this similarity, we discuss how to solve the equation for non-global logs at finite $N_c$ in Section 4,  and present numerical results in Section 5. Finally, we examine the results and conclude in Section 6.

\section{B--JIMWLK equation}
Consider a `dipole' consisting of a quark and an antiquark located at transverse coordinates $\x$ and $\y$, respectively. The $S$--matrix of the dipole moving in the $x^-=\frac{1}{\sqrt{2}}(x^0-x^3)$ direction and  scattering off some target in the eikonal approximation is
\beq
\langle S_{\x \y} \rangle_\tau = \frac{1}{N_c} \langle \tr (U_\x U^\dagger_\y) \rangle_\tau\,,
\label{S}
\eeq
 where $U_\x$ is the Wilson line  in the fundamental representation
 \beq
 (U_\x)_{ij} = P\exp\left(i\int_{-\infty}^\infty dx^- A^+_a(x^-,\x)t^a \right)_{ij}\,, \qquad 1\le i,j\le N_c\,,
 \eeq
and  $\tau$ is the rapidity of the dipole.
 The expectation value $\langle \cdots \rangle$ is taken in the target wavefunction. In deep inelastic scattering (DIS), $\tau$ is essentially the logarithm of the Bjorken--$x$ variable
 \beq
 \tau \equiv \frac{\alpha_s}{\pi} \ln \frac{1}{x}\,.
 \eeq
 The dipole $S$--matrix (\ref{S}) (more precisely, $1-\langle S\rangle$) then measures the total cross section  of the subprocess $\gamma^*p \to q\bar{q} p\to X$ at the corresponding values of $x$ and the photon virtuality $Q\sim 1/|\x-\y|$.

The B--JIMWLK\footnote{Acronym for Balitsky, Jalilian-Marian, Iancu, McLerran, Weigert, Leonidov and Kovner.} equation \cite{Balitsky:1995ub,JalilianMarian:1997gr,Iancu:2000hn} resums the small--$x$ logarithms  $\tau^n\sim (\alpha_s \ln 1/x)^n$ which arise in the high energy  evolution of $\langle S\rangle$. It  reads
\beq
\partial_\tau \langle S_{\x\y}\rangle_\tau = N_c \int \frac{d^2\z}{2\pi} {\mathcal M}_{\x\y}(\z)
\Bigl( \langle
  S_{\x\z} S_{\z\y}\rangle_\tau  -\langle S_{\x\y}\rangle_\tau \Bigr)\,,
 \label{JIM}
\eeq
 where the dipole kernel is given by
 \beq
 {\mathcal M}_{\x\y}(\z) = \frac{(\x-\y)^2}{(\x-\z)^2(\z-\y)^2}\,.  \label{f}
 \eeq
 The equation (\ref{JIM}) is not closed because the right--hand--side contains the double dipole $S$--matrix $\langle SS\rangle$. This non-linearity reflects the gluon saturation effect in the target. The consequence is that (\ref{JIM}) is actually the first equation of an infinite hierarchy of coupled equations. However, it becomes a closed equation---the Batlisky--Kovchegov (BK) equation  \cite{Balitsky:1995ub,Kovchegov:1999yj}---if one truncates the hierarchy and assumes factorization
  $\langle SS\rangle \to \langle S\rangle \langle S\rangle$
\beq
\partial_\tau \langle S_{\x\y}\rangle_\tau = N_c \int \frac{d^2\z}{2\pi} {\mathcal M}_{\x\y}(\z)
\Bigl( \langle
   S_{\x\z}\rangle_\tau \langle S_{\z\y}\rangle_\tau -\langle S_{\x\y}\rangle_\tau \Bigr)\,.
 \label{BK}
\eeq

While the BK equation (\ref{BK}) can be solved numerically in a straightforward manner,  solving the B--JIMWLK equation (\ref{JIM}) had been difficult until the ingenious reformulation of the problem as random walk \cite{Blaizot:2002np}. To explain this, it is essential to write the equation in the operator form
  \beq
\partial_\tau \langle S_{\x\y}\rangle_\tau = -\langle \hat{H} S_{\x\y} \rangle_\tau\,.  \label{fun}
\eeq
The effective Hamiltonian $\hat{H}$ takes the form
\beq
\hat{H} &=&  \int d^2\x d^2\y  \frac{d^2\z}{2\pi} {\mathcal K}_{\x\y}(\z)  \nabla^a_\x \left(1 + \tilde{U}_\x^\dagger \tilde{U}_\y -  \tilde{U}^\dagger_\x \tilde{U}_\z - \tilde{U}_\z^\dagger \tilde{U}_\y \right)^{ab}\nabla^b_\y \nonumber \\
&=& \int d^2\x d^2\y  \frac{d^2\z}{2\pi} {\mathcal K}_{\x\y}(\z)   (1 -  \tilde{U}^\dagger_\x \tilde{U}_\z )^{ac}(1- \tilde{U}_\z^\dagger \tilde{U}_\y)^{cb}\nabla^a_\x \nabla^b_\y -\int d^2\x \, \sigma_\x ^a\nabla_\x^a\,,   \label{gauge}
\eeq
  where $\U^{ab} = (\U^\dagger)^{ba}$  is the Wilson line in the adjoint representation and ${\mathcal K}$ is the gluon emission kernel
  \beq
  {\mathcal K}_{\x\y}(\z) = \frac{(\x-\z)\cdot (\z-\y)}{(\x-\z)^2 (\z-\y)^2}\,.  \label{kappa}
  \eeq
   In the last  line of (\ref{gauge}), we defined
  \beq
  \sigma_\x^a = -i\int \frac{d^2\z}{2\pi} \frac{1}{(\x-\z)^2} \tr (T^a \tilde{U}^\dagger_\x \tilde{U}_\z )\,.
  \eeq
   The derivative $\nabla$ acts on the  Wilson line in the fundamental representation as
  \beq
  \nabla^a_\x U_\y = iU_\x t^a \delta^{(2)}(\x-\y)\,, \qquad
  \nabla^a_\x U_\y^\dagger = -it^a U_\x^\dagger \delta^{(2)}(\x-\y)\,,
  \eeq
 and also on the adjoint Wilson line  with $(T^a)^{bc}=-if^{abc}$
  \beq
  \nabla^a_\x \tilde{U}_\y = i\tilde{U}_\x T^a \delta^{(2)}(\x-\y)\,, \qquad
   \nabla^a_\x \tilde{U}_\y^\dagger = -iT^a U_\x^\dagger \delta^{(2)}(\x-\y)\,.
  \label{adj}
  \eeq
The latter operation has been used in obtaining the $\sigma$--term in (\ref{gauge}).
An important observation is that (\ref{fun}) may be viewed as the Fokker--Planck equation treating  Wilson lines $U$ as dynamical variables. As is well--known, there is an associated Langevin equation which describes the random walk of these variables. The latter can be simulated numerically on a lattice, and  the solution of the B--JIMWLK equation has thus been obtained in \cite{Rummukainen:2003ns}. We shall later discuss this approach in detail. \\

Before leaving this section, it is useful to show   another form of the Hamiltonian derived in \cite{Hatta:2005as}
\beq
 \hat{H}&=&  \frac{1}{2}\int d^2\x d^2\y \frac{d^2\z}{2\pi} {\mathcal M}_{\x\y}(\z)  \nabla^a_\x\left(1 + \tilde{U}_\x^\dagger \tilde{U}_\y -  \tilde{U}^\dagger_\x \tilde{U}_\z - \tilde{U}_\z^\dagger \tilde{U}_\y  \right)^{ab} \nabla^b_\y   \nonumber \\
 &=&
  \frac{1}{2}\int d^2\x d^2\y \frac{d^2\z}{2\pi} {\mathcal M}_{\x\y}(\z)  \left(1 + \tilde{U}_\x^\dagger \tilde{U}_\y -  \tilde{U}^\dagger_\x \tilde{U}_\z - \tilde{U}_\z^\dagger \tilde{U}_\y  \right)^{ab}\nabla^a_\x \nabla^b_\y\,. \label{di}
 \eeq
  This Hamiltonian can be used only when it acts on `gauge invariant' operators  of the form
  \beq
  \tr (U_\x U^\dagger_\y)\,,   \quad \tr(U_\x U^\dagger_\y U_\z U^\dagger_{\bm w} \cdots)\,, \quad \tr(U_\x U^\dagger_\y) \,\tr(U_\z U^\dagger_{\bm w})\,, \quad \cdots
  \eeq
   and generates the same  equations as those obtained from (\ref{gauge}).
  Eq.~(\ref{di}) features the dipole kernel (\ref{f}) instead of the gluon emission kernel (\ref{kappa}). The following relation between the two kernels is worth noting
  \beq
  {\mathcal M}_{\x\y}(\z) = 2{\mathcal K}_{\x\y}(\z) -{\mathcal K}_{\x\x}(\z) -{\mathcal K}_{\y\y}(\z)\,.  \label{rel}
  \eeq

\section{Non-global logs at finite $N_c$}

\begin{figure}[t]
\begin{center}
\includegraphics[height=8cm]{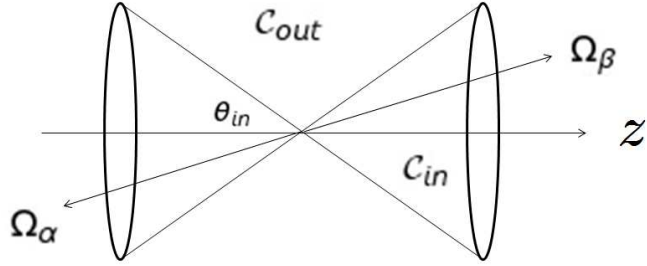}
\caption{Back--to--back jets. \label{1} }
\end{center}
\end{figure}

We now turn to the resummation of non-global logarithms which is our primary interest. Consider, as in the final state of $e^+e^-$ annihilation, a pair of jets   that is overall color singlet and pointing in the direction
 of the solid angle  $\Omega_{\alpha,\beta}=(\theta_{\alpha,\beta},\phi_{\alpha,\beta})$ measured with respect to the positive $z$--axis (see Fig.~\ref{1}). Let ${\mathcal C}_{in}$ be the region inside a pair of back--to--back cones with opening angle $\theta_{in}$ which include the jets, and let ${\mathcal C}_{out}$ be its complementary region. We then ask what is the probability $P(\Omega_\alpha,\Omega_\beta)$ that the total flow of energy into ${\mathcal C}_{out}$ is less than $E_{out}$. In the perturbative calculation of $P$ in the regime $Q\gg E_{out}\gg \Lambda_{QCD}$ where $Q$ is the hard scale, logarithms of the form $(\alpha_s \ln Q/E_{out})^n$ appear which have to be resummed. These logarithms are non-global because the measurement is done only in ${\mathcal C}_{out}$. To leading logarithmic accuracy, one may identify $E_{out}$ with the jet veto scale $p_T^{\rm veto}$ mentioned in the introduction.

It has been shown by Banfi, Marchesini and Smye (BMS) that $P$ satisfies the following evolution equation \cite{Banfi:2002hw}
 \beq
 \partial_\tau  P_{\alpha\beta}= N_c \int \frac{d\Omega_\gamma}{4\pi}
 {\mathcal M}_{\alpha\beta}(\gamma)
\Bigl( \Theta_{in}(\gamma) P_{\alpha \gamma} P_{\gamma\beta}
-P_{\alpha\beta} \Bigr)
\,,   \label{bms}
\eeq
where we abbreviated $P_{\alpha\beta}=P(\Omega_\alpha,\Omega_\beta)$ and the  parameter $\tau$ is defined as
\beq
 \tau = \begin{cases} \frac{\alpha_s}{\pi} \ln \frac{Q}{E_{out}}\,, & (\mbox{fixed coupling}) \\
 \frac{6}{11N_c-2n_f} \ln \left(\frac{\ln Q/\Lambda_{QCD}}{\ln E_{out}/\Lambda_{QCD}}
 \right)\,. & (\mbox{running coupling})
 \end{cases}
 \label{tau}
 \eeq
  The equation (\ref{bms}) resums logarithms $\tau^n\sim (\alpha_s \ln E_{out})^n$ to all orders.
  The integral kernel
 \beq
 {\mathcal M}_{\alpha\beta}(\gamma) &\equiv & \frac{1-\cos \theta_{\alpha\beta}}{(1-\cos \theta_{\alpha \gamma})(1-\cos \theta_{\gamma\beta})} \nonumber \\  &=& \frac{1-\n_\alpha \cdot \n_\beta}{(1-\n_\alpha \cdot \n_\gamma)(1-\n_\gamma\cdot \n_\beta)} \nonumber \\
 &=& \frac{n_\alpha \cdot n_\beta}{(n_\alpha \cdot n_\gamma) (n_\gamma \cdot n_\beta) }\,,  \label{sug}
  \eeq
    is composed of null vectors $n^\mu = (1,\sin \theta \cos \phi, \sin \theta \sin \phi, \cos \theta)=(1,\n)$  proportional to the four--vector of hard partons.
The `step function'
 \beq
 \Theta_{in}(\gamma) \equiv \begin{cases} 1 & (\Omega_\gamma \in {\mathcal C}_{in})\,, \\
  0 & (\Omega_\gamma \in {\mathcal C}_{out})\,,
  \end{cases}
  \eeq
  ensures that real gluons are emitted only in ${\mathcal C}_{in}$.\footnote{ In principle, real gluons can be directly emitted from hard partons into ${\mathcal C}_{out}$ provided their energy is less than $E_{out}$. However, to leading logarithmic accuracy such contributions may be omitted \cite{Banfi:2002hw}.}  A trivial rewriting of (\ref{bms})
   \beq
 \partial_\tau  P_{\alpha\beta} =  -N_c \int \frac{d\Omega_\gamma}{4\pi}
 {\mathcal M}_{\alpha\beta}(\gamma) \Theta_{out}(\gamma)
P_{\alpha\beta} + N_c \int \frac{d\Omega_\gamma}{4\pi}
 {\mathcal M}_{\alpha\beta}(\gamma) \Theta_{in}(\gamma) \Bigl( P_{\alpha \gamma} P_{\gamma\beta}
-P_{\alpha\beta} \Bigr)
\,,    \nonumber \\
\eeq
where  $\Theta_{out}(\gamma) \equiv 1-\Theta_{in}(\gamma)$
   illuminates the physical meaning of  the right--hand--side. The first term represents the familiar Sudakov suppression, whereas non-global logarithms are resummed by the second term describing the emission of an arbitrary number of gluons into ${\mathcal C}_{in}$ which then coherently emit the softest gluons into  ${\mathcal C}_{out}$.

  For the purpose of the present work, it is indispensable  to note that the equation (\ref{bms}) has been derived in the large--$N_c$ limit. Its finite--$N_c$ generalization was discussed by Weigert \cite{Weigert:2003mm} and the result reads
\beq
\partial_\tau \langle P_{\alpha\beta}\rangle_\tau
&=& N_c \int \frac{d\Omega_\gamma}{4\pi} {\mathcal M}_{\alpha\beta}(\gamma)
\left\langle
\Theta_{in}(\gamma) \left(P_{\alpha \gamma} P_{\gamma\beta} -\frac{P_{\alpha\beta}}{N_c^2}\right)
-\frac{2C_F}{N_c} P_{\alpha\beta}
\right\rangle_\tau
\nonumber \\
&=&
-2C_F \int \frac{d\Omega_\gamma}{4\pi} {\mathcal M}_{\alpha\beta}(\gamma)
\Theta_{out}(\gamma) \langle P_{\alpha\beta}\rangle_\tau \nonumber \\
&& \qquad \qquad \quad +
N_c \int \frac{d\Omega_\gamma}{4\pi} {\mathcal M}_{\alpha\beta}(\gamma)
\Theta_{in}(\gamma) \big\langle P_{\alpha \gamma} P_{\gamma\beta} -P_{\alpha\beta}\big\rangle_\tau
\,,  \label{finite}
\eeq
 where $C_F=\frac{N_c^2-1}{2N_c}$. However, (\ref{finite}) itself is highly formal in that the meaning of the averaging $\langle \cdots \rangle$  can be specified only indirectly, as a proxy of certain complicated functional integrals \cite{Weigert:2003mm}.
 Nevertheless, putting this qualification aside,  the striking similarity between Eqs.~(\ref{bms}), (\ref{finite}) and Eqs.~(\ref{JIM}), (\ref{BK})  is unmistakable.  In fact, it is possible to establish a rigorous mathematical equivalence between the two problems. As shown in  \cite{Hatta:2008st,Avsar:2009yb}, a conformal transformation known as the stereographic projection
\beq
\x=(x^1,x^2) =\left(\frac{\sin \theta \cos \phi}{1+\cos \theta}, \frac{\sin \theta \sin \phi}{1+\cos \theta}\right)\,,
\eeq
exactly maps the respective kernels onto each other
\beq
\frac{d^2\z}{2\pi} \frac{(\x-\y)^2}{(\x-\z)^2(\z-\y)^2} = \frac{d\Omega_\gamma}{4\pi}
\frac{1-\cos \theta_{\alpha\beta}}{(1-\cos \theta_{\alpha \gamma})(1-\cos \theta_{\gamma\beta})}\,.
\label{cru}
\eeq
Aside from the kinematical constraint factor $\Theta_{in}$, the map (\ref{cru}) dispels any structural difference between the two sets of equations.\footnote{Note that if one sets $\Theta_{in}\to 1$, (\ref{finite}) becomes  identical to (\ref{JIM}) under the map (\ref{cru}).} This equivalence probably has a deep geometrical origin which goes beyond the perturbative framework. Indeed, such a correspondence  persists even in the strong coupling limit of ${\mathcal N}=4$ supersymmetric Yang--Mills theory \cite{Hatta:2008st}.

  Fortunately, a powerful machinery to solve the JIMWLK  problem (\ref{JIM}) is available, and this
 brings hope that the jet problem (\ref{finite}) can be solved in a similar manner.  For this purpose, one seeks the operator form of (\ref{finite}) with a formal identification
 \beq
 P_{\alpha\beta} \leftrightarrow \frac{1}{N_c} \tr (U_\alpha U^\dagger_\beta)\,.
 \label{id}
 \eeq
  Here the Wilson lines $U_{\alpha,\beta}$ represent eikonal (jet) lines  starting from the space-time origin and extending to infinity in the direction of $\Omega_{\alpha,\beta}$.\footnote{Given the probabilistic nature of $P$, one should more properly consider $U$ as the product of a Wilson line in the amplitude and that in the complex--conjugate amplitude \cite{Weigert:2003mm}. In practice, however, this does not matter in the equivalent Langevin approach.}  Eq.~(\ref{finite}) can then be written as a Fokker--Planck equation in group space
 \beq
\partial_\tau \langle P_{\alpha\beta}\rangle_\tau =- \langle \hat{H} P_{\alpha\beta} \rangle\,,
\label{gahaha}
\eeq
where \cite{Weigert:2003mm}
\beq
\hat{H} &=&  \frac{1}{2}\int d\Omega_\alpha d\Omega_\beta \frac{d\Omega_\gamma}{4\pi} {\mathcal M}_{\alpha\beta}(\gamma)  \nabla^a_\alpha \left(1 + \tilde{U}_\alpha^\dagger \tilde{U}_\beta -  \Theta_{in}(\gamma) \bigl(\tilde{U}^\dagger_\alpha \tilde{U}_\gamma + \tilde{U}_\gamma^\dagger \tilde{U}_\beta \bigr) \right)^{ab}\nabla^b_\beta \nonumber \\
&=& \frac{1}{2}\int d\Omega_\alpha d\Omega_\beta \frac{d\Omega_\gamma}{4\pi} {\mathcal M}_{\alpha\beta}(\gamma)  \left(1 + \tilde{U}_\alpha^\dagger \tilde{U}_\beta -  \Theta_{in}(\gamma) \bigl(\tilde{U}^\dagger_\alpha \tilde{U}_\gamma + \tilde{U}_\gamma^\dagger \tilde{U}_\beta \bigr) \right)^{ab}\nabla^a_\alpha \nabla^b_\beta\,.  \label{bm}
\eeq
 In (\ref{bm}),
 the derivative $\nabla$ is defined by ($\delta (\Omega -\Omega') \equiv \delta (\cos \theta -\cos \theta') \delta (\phi-\phi'))$
  \beq
  \nabla^a_\alpha U_\beta = iU_\alpha t^a \delta(\Omega_\alpha- \Omega_\beta)\,, \qquad
  \nabla^a_\alpha U_\beta^\dagger = -it^a U_\alpha^\dagger \delta (\Omega_\alpha-\Omega_\beta)\,, \label{nor}
  \eeq
  and similarly to (\ref{adj}) in the adjoint case ($t^a \to T^a$).
 \\

\section{Equivalent Langevin dynamics}

The Fokker--Planck equations (\ref{fun}) and (\ref{gahaha}) can be solved by making use of  an equivalent Langevin formulation.
 To illustrate the  idea, consider the following Fokker--Planck equation  for some probability distribution $P_\tau(x)$ of dynamical variables $\{x^a\}$
\beq
\frac{dP_\tau(x)}{d\tau} &=&  \frac{1}{2} \partial_a \left(\chi^{ab}(x)\partial_b P_\tau(x) \right) \\
&=& \left(\frac{1}{2} \chi^{ab}(x)\partial_a \partial_b +\sigma^a \partial_a \right)P_\tau(x) \\ &=&  \partial_a \left[ \frac{1}{2}  \partial_b \left( \chi^{ab}(x) P_\tau(x)\right) - \sigma^a  P_\tau(x) \right]
\,, \label{fp}
\eeq
 where $\chi^{ab}=\chi^{ba}$ and $\sigma^a \equiv \frac{1}{2}\partial_b \chi^{ba}$.
 We assume that $\chi$ is factorized in the form $\chi^{ab} = {\mathcal E}^{ac}{\mathcal E}^{cb}$.
The equivalent Langevin equation is then
\beq
\frac{d}{d\tau} x^a(\tau) = \sigma^a (x) + {\mathcal E}^{ac}(x)\xi_c(\tau)\,, \label{lan}
\eeq
 where $\xi$ is the Gaussian white noise characterized by the correlator
  \beq
\langle \xi_a(\tau)\xi_b(\tau') \rangle = \delta_{ab} \delta(\tau-\tau')\,.
\eeq
The distribution can then be obtained by averaging over an ensemble $\{x^a(\tau)\}$ of random walk trajectories
\beq
P_\tau(x) = \langle \delta(x-x(\tau))\rangle\,.
\eeq

 To be more precise, the equation (\ref{lan})
 makes sense only in a $\tau$--discretized form. There is a well--known ambiguity in how we discretize the equation, the so--called It$\hat{\mbox{o}}$--Stratonovich dilemma. The appropriate choice  corresponding to  (\ref{fp}) is the It$\hat{\mbox{o}}$ scheme
 \beq
 x^a(\tau + \varepsilon) = x^a(\tau) + \varepsilon \sigma^a(x) + \sqrt{\varepsilon}{\mathcal E}^{ac}(x(\tau))\xi_c(\tau)\,, \label{ito}
 \eeq
 where $\varepsilon$ is the time step and the argument of ${\mathcal E}$ is evaluated at the previous time $\tau$, respecting causality \cite{zinn}. In (\ref{ito}), we have rescaled the noise  as $\sqrt{\varepsilon}\xi \to \xi$ in order to make explicit the fact that the typical variation $\Delta x$ is ${\mathcal O}(\sqrt{\varepsilon})$ in a random walk. With this normalization, the noise correlator in discrete time reads
 \beq
 \langle \xi_a(\tau)\xi_b(\tau')\rangle =\delta_{ab}\delta_{\tau\tau'}\,. \label{norm}
 \eeq

The operator form of the B--JIMWLK equation (\ref{fun}) has precisely the structure (\ref{fp}) with a factorized kernel (\ref{gauge}).
  It can thus be described by the Langevin dynamics (\ref{ito}), with the SU(3) matrices $U_{\x,\y}$ playing the role of  $\{x^a\}$
 \cite{Blaizot:2002np}.  One can simulate the random walk in group space on a transverse lattice, and this is how the solution to B--JIMWLK equation has been originally obtained \cite{Rummukainen:2003ns}.

  In \cite{Weigert:2003mm}, Weigert suggested to follow the same strategy in solving (\ref{gahaha}). A comparison of  (\ref{bm})  and (\ref{fp}) implies that $\sigma=0$. The kernel of (\ref{bm})
  \beq
 && 1 + \tilde{U}_\alpha^\dagger \tilde{U}_\beta -  \Theta_{in}(\gamma) \bigl(\tilde{U}^\dagger_\alpha \tilde{U}_\gamma + \tilde{U}_\gamma^\dagger \tilde{U}_\beta \bigr) \nonumber \\
  && \qquad \qquad = \Theta_{out}(\gamma)+\Theta_{out}(\gamma) \tilde{U}^\dagger_\alpha \tilde{U}_\beta + \Theta_{in} (\gamma) (1-\tilde{U}_\alpha^\dagger \tilde{U}_\gamma)(1-\tilde{U}^\dagger_\gamma \tilde{U}_\beta)\,, \label{w}
  \eeq
    can be written as a sum of three factorized terms. [Note that $(\Theta_{in})^2=\Theta_{in}$.]  Exploiting the factorized form of ${\mathcal M}$  as seen in the last line of (\ref{sug}), Weigert deduced  an analog of the stochastic term ${\mathcal E}^{ac}\xi_c$ in (\ref{lan}) by introducing three independent noises $\xi^{(I)}$, ($I=1,2,3$)
  \beq
  {\mathcal E}^{ac}\xi_c \sim  \int d\Omega_\gamma \frac{n^\mu_\alpha}{n_\alpha \cdot n_\gamma} \left\{ \Theta_{out}(\gamma) \left(\delta^{ac} \xi^{(1)}_{\gamma c\mu}+ (\tilde{U}^\dagger_\alpha)^{ac} \xi^{(2)}_{\gamma c\mu} \right)+ \Theta_{in}(\gamma) (1-\tilde{U}^\dagger_\alpha \tilde{U}_\gamma)^{ac}\xi^{(3)}_{\gamma c\mu} \right\}\,, \nonumber \\
   \eeq
    which generates a random walk $U\to Ue^{it^a{\mathcal E}^{ac}\xi_c}$ in group space.
However, the fact that ${\mathcal M}$ is factorized in four--vector space means that the  noises must have the correlator
\beq
\langle \xi_a^{(I)\mu} \xi_b^{(J)\nu}\rangle \sim \delta_{ab}\delta^{IJ}g^{\mu\nu}\,,
\eeq
which is negative for the spatial components $\mu,\nu =1,2,3$, hence they cannot be simulated in practice.

The resolution of this problem again comes from the correspondence with the B--JIMWLK evolution. Firstly, the identity  (\ref{cru}) clearly shows the two--dimensional nature of the dipole kernel, so introducing four--vectors is an excess. We then notice the close similarity  between (\ref{bm}) and (\ref{di}), the latter being equivalent to (\ref{gauge}). This suggests that we can rewrite the effective Hamiltonian in a form analogous to (\ref{gauge}) also for the jet problem.
 We thus look for a kernel ${\mathcal K}$ satisfying (cf. (\ref{rel}))
\beq
{\mathcal M}_{\alpha\beta}(\gamma) = 2{\mathcal K}_{\alpha\beta}(\gamma) -{\mathcal K}_{\alpha\alpha}(\gamma)-{\mathcal K}_{\beta\beta}(\gamma)\,. \label{look}
\eeq
The solution is found to be (cf. (\ref{kappa}))
\beq
{\mathcal K}_{\alpha\beta}(\gamma) =  \frac{\cos \theta_{\alpha\gamma} + \cos \theta_{\gamma\beta}-\cos \theta_{\alpha\beta} -1}{2(1-\cos\theta_{\alpha\gamma})(1-\cos \theta_{\gamma\beta})} =\frac{(\n_\alpha -\n_\gamma)\cdot (\n_\gamma- \n_\beta)}{2(1-\n_\alpha \cdot \n_\gamma)(1-\n_\gamma \cdot \n_\beta)}\,,
\eeq
which has a factorized structure on the unit sphere ($|\n|=1$) embedded in three spatial dimensions.
 Reversing the argument in \cite{Hatta:2005as} which led (\ref{gauge}) to (\ref{di}), we arrive at an effective Hamiltonian equivalent to (\ref{bm})
 \beq
 \hat{H} &=&  \int d\Omega_\alpha d\Omega_\beta \frac{d\Omega_\gamma}{4\pi} {\mathcal K}_{\alpha\beta}(\gamma) \nabla^a_\alpha \left(1 + \tilde{U}_\alpha^\dagger \tilde{U}_\beta -  \Theta_{in}(\gamma) \bigl(\tilde{U}^\dagger_\alpha \tilde{U}_\gamma + \tilde{U}_\gamma^\dagger \tilde{U}_\beta \bigr) \right)^{ab} \nabla^b_\beta \nonumber \\
 &=& \int d\Omega_\alpha d\Omega_\beta \frac{d\Omega_\gamma}{4\pi} {\mathcal K}_{\alpha\beta}(\gamma)  \left( 1 + \tilde{U}_\alpha^\dagger \tilde{U}_\beta -  \Theta_{in}(\gamma) \bigl(\tilde{U}^\dagger_\alpha \tilde{U}_\gamma + \tilde{U}_\gamma^\dagger \tilde{U}_\beta \bigr)  \right)^{ab}\nabla^a_\alpha \nabla^b_\beta
 \nonumber \\ &&  \quad
  -\int d\Omega_\alpha \sigma_\alpha^a \nabla^a_\alpha\,,
 \eeq
 where
 \beq
 \sigma_\alpha^a &=& -i\int \frac{d\Omega_\gamma}{4\pi} \frac{\Theta_{in}(\gamma)}{(1-\n_\alpha \cdot \n_\gamma)}
 \tr (T^a \U^\dagger_\alpha \U_\gamma)\,.
 \eeq
 Moreover, differently from (\ref{w}), we write the expression in the brackets  as a sum of two factorized terms
 \beq
 && 1 + \tilde{U}_\alpha^\dagger \tilde{U}_\beta -  \Theta_{in}(\gamma) \bigl(\tilde{U}^\dagger_\alpha \tilde{U}_\gamma + \tilde{U}_\gamma^\dagger \tilde{U}_\beta \bigr)\nonumber \\
 && \qquad \quad  =
 (1 -  \Theta_{in}(\gamma) \tilde{U}^\dagger_\alpha \tilde{U}_\gamma)(1- \Theta_{in} \tilde{U}_\gamma^\dagger \tilde{U}_\beta ) + \Theta_{out}(\gamma) \tilde{U}_\alpha^\dagger \tilde{U}_\beta\,.
 \eeq
 This reduces the number of independent noises from three to two $\xi^{(I)}$ ($I=1,2$). They are   characterized by the following correlator (in discretized `time' $\tau$, cf. (\ref{norm}))
 \beq
 \langle \xi^{(I)k}_{\alpha a}  (\tau) \xi^{(J)l}_{\beta b} (\tau') \rangle = \delta_{\tau\tau'} \delta (\cos \theta_\alpha -\cos \theta_\beta)\delta(\phi_\alpha- \phi_\beta) \delta^{IJ} \delta_{ab}\delta^{kl}\,, \label{true}
 \eeq
  where $k,l=1,2,3$ are the spatial indices. The problem of negative metric has been circumvented.
 We can now write down the associated Langevin evolution for $U_\alpha$ with $\Omega_\alpha \in {\mathcal C}_{in}$ (cf. (\ref{ito}))
 \beq
 U_\alpha (\tau+\varepsilon) =U_\alpha (\tau) \exp\left\{ it^a \left( \sqrt{\varepsilon} \int d\Omega_\gamma \, \left({\bm {\mathcal E}}_{\alpha\gamma}^{(1)ac} \cdot {\bm \xi}_{\gamma c}^{(1)}  +{\bm {\mathcal E}}_{\alpha\gamma}^{(2)ac} \cdot {\bm \xi}_{\gamma c}^{(2)} \right) +\varepsilon \sigma_\alpha^a \right) \right\}\,, \label{right}
 \eeq
  where
 \beq
 ({\bm {\mathcal E}}_{\alpha\gamma}^{(1)ac})^k = \frac{1}{\sqrt{4\pi}} \frac{(\n_\alpha-\n_\gamma)^k}{1-\n_\alpha \cdot \n_\gamma} [1-\Theta_{in}(\gamma)\tilde{U}^\dagger_\alpha \tilde{U}_\gamma ]^{ac}\,,
\eeq
\beq
({\bm {\mathcal E}}_{\alpha\gamma}^{(2)ac})^k = \frac{1}{\sqrt{4\pi}} \frac{(\n_\alpha-\n_\gamma)^k}{1-\n_\alpha\cdot \n_\gamma} \Theta_{out}(\gamma) (\tilde{U}_\alpha^\dagger)^{ac}\,.
\eeq
 All the matrices on the right--hand--side of (\ref{right}) are evaluated at $\tau$ according to the It$\hat{\mbox{o}}$ scheme. Expanding the exponential up to ${\mathcal O}(\varepsilon)$, we get
 \beq
U_\alpha(\tau+\varepsilon) &\approx & U_\alpha(\tau)
  +i\sqrt{\varepsilon}U_\alpha (\tau)t^a   \int d\Omega_\gamma \, \left({\bm {\mathcal E}}_{\alpha\gamma}^{(1)ac} \cdot {\bm \xi}_{\gamma c}^{(1)}  +{\bm {\mathcal E}}_{\alpha\gamma}^{(2)ac} \cdot {\bm \xi}_{\gamma c}^{(2)} \right) \nonumber \\
   && +\varepsilon U_\alpha (\tau) \left\{ it^a   \sigma_\alpha^a -\frac{1}{2}
  \left( t^a \int d\Omega_\gamma \, \left({\bm {\mathcal E}}_{\alpha\gamma}^{(1)ac} \cdot {\bm \xi}_{\gamma c}^{(1)}  +{\bm {\mathcal E}}_{\alpha\gamma}^{(2)ac} \cdot {\bm \xi}_{\gamma c}^{(2)} \right) \right)^2
   \right\} \,. \label{fin}
\eeq
The second line of (\ref{fin}) can be simplified as follows. In the $\sigma$--term, we use the identities $\U^{ab}t^b = U^\dagger t^a U$ and $t^a Ut^a = \frac{1}{2}(\tr U)  -\frac{1}{2N_c}U$ to obtain
   \beq
  U_\alpha t^a \, \tr (T^a \U_\alpha^\dagger \U_\gamma) &=& -U_\alpha [t^b,t^c](\U_\alpha^\dagger)^{cd}(\U_\gamma)^{db} \nonumber \\
  &=& \frac{1}{2}\tr (U_\alpha U^\dagger_\gamma)\, U_\gamma -\frac{1}{2} \tr (U_\gamma U^\dagger_\alpha)\, U_\alpha U_\gamma^\dagger U_\alpha\,.
  \eeq
  To the accuracy of ${\mathcal O}(\varepsilon)$, the terms quadratic in noise $\xi \xi$ may be replaced by their expectation values using (\ref{true}). After these manipulations,  (\ref{fin}) takes the form
 \beq
&& U_\alpha(\tau+\varepsilon) = U_\alpha(\tau) \nonumber \\
  && \ \ + i\sqrt{\frac{\varepsilon}{4\pi}} \int d\Omega_\gamma  \frac{(\n_\alpha-\n_\gamma)^k}{1-\n_\alpha\cdot \n_\gamma} \biggl( U_\alpha t^a \xi^{(1)k}_{\gamma a}  -\Theta_{in}(\gamma) U_\gamma t^a U_\gamma^\dagger U_\alpha \xi^{(1)k}_{\gamma a} +\Theta_{out}(\gamma)t^a U_\alpha \xi^{(2)k}_{\gamma a} \biggr) \nonumber \\
 &&\ \ +\varepsilon  \int \frac{d\Omega_\gamma}{4\pi} \frac{1}{1-\n_\alpha\cdot \n_\gamma}
 \left(-2C_F U_\alpha + \Theta_{in}(\gamma) \left(\tr (U_\alpha U_\gamma^\dagger)\, U_\gamma -\frac{1}{N_c} U_\alpha\right) \right)
 \,.  \label{final}
 \eeq
  Note that there is no singularity at
 $\Omega_\gamma = \Omega_\alpha \in {\mathcal C}_{in}$.
 By computing the difference \beq
 \frac{1}{N_c} \tr(U_\alpha(\tau+\varepsilon) U^\dagger_\beta (\tau+\varepsilon)) - \frac{1}{N_c} \tr(U_\alpha (\tau) U^\dagger_\beta(\tau))\,,
  \eeq
  to ${\mathcal O}(\varepsilon)$ and using (\ref{true}),  (\ref{look}) and the relation ${\mathcal K}_{\alpha\alpha}(\gamma)=-1/(1-\n_\alpha\cdot \n_\gamma)$,  one can recover  (\ref{finite}) after the identification (\ref{id}).

 For the sake of numerical simulations, it is more economical  to express the evolution (\ref{final}) in a  left--right symmetric form\footnote{See Ref.~\cite{Lappi:2012vw} for a similar rewriting of the JIMWLK evolution. As noted in this paper, the right--multiplication rule  such as (\ref{right}) is related to  the choice of the Hamiltonian (\ref{gauge}) or (\ref{bm}) being expressed in terms of `right--derivatives' $\nabla^a_R U\sim Ut^a$. The $\sigma$--term can be eliminated in the process of converting one of the  right--derivatives into a left--derivative $\nabla_L^a U\sim t^a U$.}
\beq
U_\alpha(\tau+\varepsilon) &=& e^{iA^L_\alpha} U_\alpha (\tau) e^{iA^R_\alpha}\,, \label{alt}
\eeq
 where
 \beq
 A^L_\alpha = \sqrt{\frac{\varepsilon}{4\pi}}\int d\Omega_\gamma \frac{(\n_\alpha -\n_\gamma)^k}{1-\n_\alpha \cdot \n_\gamma} \left( -\Theta_{in}(\gamma) U_\gamma t^a U_\gamma^\dagger  \xi^{(1)k}_{\gamma a} +\Theta_{out}(\gamma)\, t^a  \xi^{(2)k}_{\gamma a} \right)\,,
 \eeq
 \beq
 A^R_\alpha= \sqrt{\frac{\varepsilon}{4\pi}} \int d\Omega_\gamma  \frac{(\n_\alpha -\n_\gamma)^k}{1-\n_\alpha \cdot \n_\gamma} \, t^a \xi^{(1)k}_{\gamma a}\,.
 \eeq
 In this representation only the terms proportional to noise are kept in the exponential.
 It is easy to check that (\ref{alt}) and (\ref{final}) are equivalent to ${\mathcal O}(\varepsilon)$ under the identification $\xi\xi \approx \langle \xi \xi\rangle$.
 Eqs.~(\ref{true}) and (\ref{alt}) will serve as the starting point of our numerical simulation.

\section{Numerical simulation}

 We simulate the random walk (\ref{alt}) of Wilson lines living on the unit sphere by discretizing the coordinates $1\ge \cos \theta\ge -1$ and $2\pi>  \phi\ge 0$ with lattice spacings $a_c$ and $a_\phi$, respectively.  The SU(3) matrices $U_\alpha$ are defined only at the grid points belonging to ${\mathcal C}_{in}$, whereas the noises $\xi_\alpha^{(I)}$ are defined at all grid points.\footnote{At $\cos \theta_\alpha=\pm 1$, $U_\alpha$ and $\xi_\alpha$ are  independent of $\phi$, as they should. In the case of $U_\alpha$, this is guaranteed by our initial condition (\ref{initial}) and the structure of the evolution (\ref{final}) which preserves this property. } The initial condition at $\tau=0$ is simply given by
\beq
U_\alpha=1\,, \ \  \mbox{(unit matrix)} \label{initial}
\eeq
for all $\Omega_\alpha \in {\mathcal C}_{in}$, or equivalently, $P_{\alpha\beta}=1$ for all pairs $(\Omega_\alpha,\Omega_\beta)$ corresponding to no radiation before evolution.\footnote{In the JIMWLK case, the initial condition is the value of the $S$--matrix (\ref{S}) at small, but not too small value of $x$. This is model dependent and its initial sampling is non-trivial \cite{Rummukainen:2003ns}.}
We then update $\{U_\alpha\}$ after each time step $\varepsilon$ according to the formula (\ref{alt}) with noises ${\bm \xi}^{(I)}$ ($I=1,2$) randomly generated from the Gaussian distribution
 \beq
\prod_{\gamma,a,k}\sqrt{\frac{a_c a_\phi}{2\pi}} \exp\left(-\frac{a_c a_\phi}{2}\xi_{\gamma a}^{(I)k}\xi_{\gamma a}^{(I)k}\right)\,, \qquad \langle \xi_{\alpha a}^{(I)k} \xi_{\beta b}^{(J)l}\rangle = \frac{1}{a_c a_\phi}\delta^{IJ}\delta_{ab}\delta^{kl}\delta_{\alpha\beta}\,.
\eeq
  In order to ensure that $U$'s remain unitary during the evolution, we need to evaluate the exponential of matrices $e^{iA_{L/R}}$ accurately (although the equation (\ref{alt}) makes sense only to ${\mathcal O}(\varepsilon)$).
In practice, we use an approximation $e^{iA} =(e^{iA/2^n})^{2^n} \approx \left(1+\frac{iA}{2^n} +\cdots +\frac{1}{m!}\left(\frac{iA}{2^n}\right)^m \right)^{2^n}$ with $m, n$ large enough. On top of this, we perform the `reunitarization' of $U$'s using polar decomposition method after every 100 steps of evolution. The  effect of this latter operation is actually very small due to our accurate evaluation of $e^{iA_{L/R}}$.

The above procedure is repeated a desired number of times $N=\tau/\varepsilon$, and at the end of this random walk trajectory we compute the trace
\beq
\frac{1}{N_c} \tr ( U_{\alpha}(\tau) U_{\beta}^\dagger(\tau) )\,.
\eeq
 We then average it over many such trajectories and identify the result with $\langle P_{\alpha\beta}\rangle_\tau$. In practice, we choose  $\varepsilon = 10^{-5}$ and average over 500 independent trajectories.

\begin{figure}[tb]
\begin{center}
\includegraphics[height=11cm]{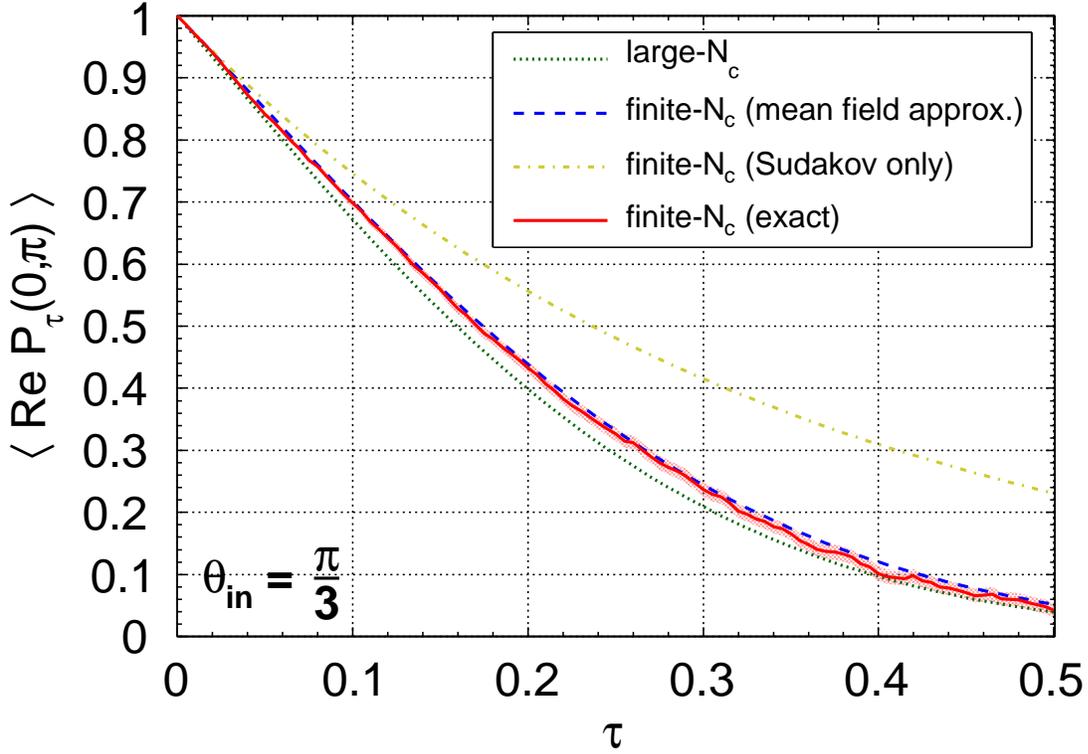}
\end{center}
\caption{ Solid line (red): exact $N_c=3$ solution to (\ref{finite}). The band indicates the standard error.  Dashed line (blue):  $N_c=3$, mean--field solution to  (\ref{finitemean}). Dotted line (green):  solution to the BMS equation (\ref{bms}) from \cite{Hatta:2009nd}.  Dash--dotted line (yellow): result with only the Sudakov term. \label{2} }
\end{figure}

Fig.~\ref{2} shows the evolution of $P_\tau (\Omega_\alpha,\Omega_\beta)$ for $\cos \theta_\alpha =1$ and $\cos \theta_{\beta}=-1$, corresponding to back--to--back jets in the beam direction.\footnote{We actually plot the real part of the average $\langle \tr U_\alpha U_\beta^\dagger \rangle$.  $\tr U_\alpha U_\beta^\dagger$ is in general complex--valued for each trajectory,  but we have checked that the imaginary part of the average  is consistent with zero within errors.}
 The opening angle of the cones (see Fig.~\ref{1}) is fixed to  $\cos \theta_{in}=\frac{1}{2}$.
The simulation was done on a lattice with 80 grid points in the $\cos \theta$ direction, and 40 grid points in the
$\phi$ direction.
The exact $N_c=3$ solution to (\ref{finite}) (solid red line) is compared with the solution of the large--$N_c$ BMS equation (\ref{bms}) (dotted green line) previously obtained in \cite{Hatta:2009nd},\footnote{Note that the definition of $\tau$ in \cite{Hatta:2009nd} differs from (\ref{tau}) by a factor of $N_c$.} and also with the solution of the `mean field approximation' to (\ref{finite}) (dashed blue line)
 \beq
\partial_\tau \langle P_{\alpha\beta}\rangle_\tau
&=&
-2C_F \int \frac{d\Omega_\gamma}{4\pi} {\mathcal M}_{\alpha\beta}(\gamma)
\Theta_{out}(\gamma) \langle P_{\alpha\beta}\rangle_\tau \nonumber \\
&&
\qquad +N_c \int \frac{d\Omega_\gamma}{4\pi} {\mathcal M}_{\alpha\beta}(\gamma)
\Theta_{in}(\gamma) \bigl(\langle P_{\alpha \gamma}\rangle_\tau \langle P_{\gamma\beta}\rangle_\tau  -\langle P_{\alpha\beta}\rangle_\tau \bigr)
\,,  \label{finitemean}
\eeq
which differs from the BMS equation only by the coefficient of the Sudakov term $N_c=3 \leftrightarrow 2C_F=8/3$.  The latter  serves as an indicator of the quality of the mean field approximation $\langle PP\rangle \to \langle P\rangle \langle P\rangle$. For the sake of reference, we also plot the solution obtained by keeping only the Sudakov term (first term on the right--hand--side) in (\ref{finitemean}) (dash--dotted yellow line).

\section{Discussion}

\begin{figure}[htb]
\begin{center}
\includegraphics[height=10cm]{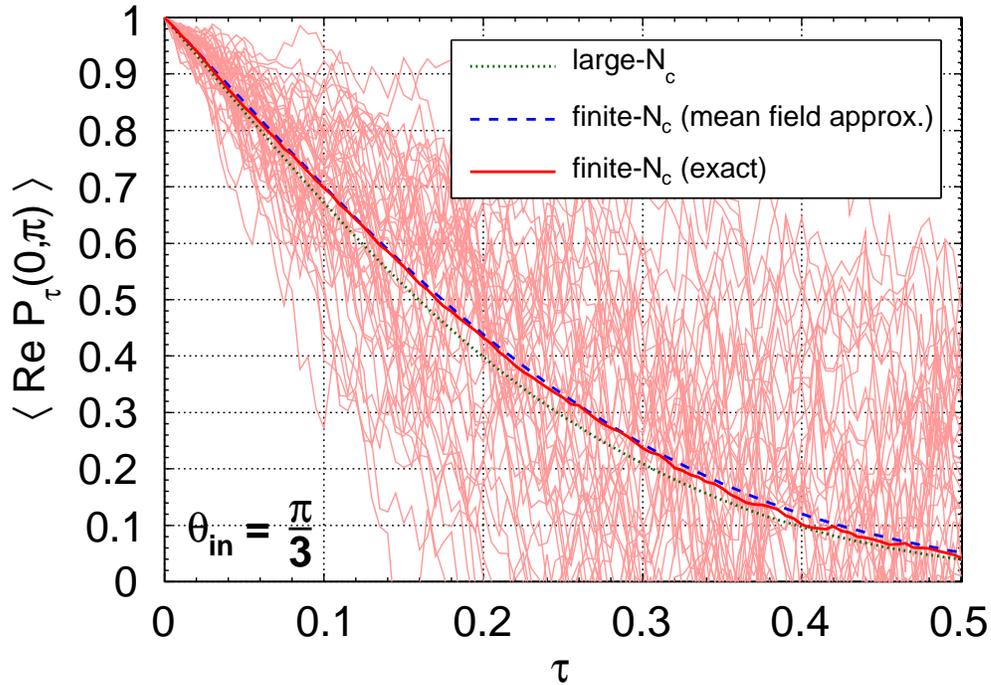}
\end{center}
\caption{Thin lines (pink): 50  random walk trajectories used in the computation of the average (thick red line).  \label{3} }
\end{figure}

 A comparison of the solid (red) and dashed (blue) lines in Fig.~\ref{2} shows that the exact solution is rather close to the mean field solution with the finite--$N_c$ corrected Sudakov term  for all values of $\tau$ explored in this work.
 This may come as no surprise to those who are acquainted with the solution of the JIMWLK  equation which agrees well with the BK solution \cite{Rummukainen:2003ns,Kovchegov:2008mk}. Nevertheless, we find the present result quite intriguing because we actually observed enormous trajectory--by--trajectory fluctuations. In Fig.~\ref{3} we show 50 (out of 500) individual random walk trajectories used in the computation of the average. It turns out that the fluctuations are so large that the standard deviation $\delta P\equiv \sqrt{\langle PP\rangle -\langle P\rangle\langle P\rangle}$ is of the order of $\langle P\rangle$ itself for not--so--small values of $\tau$. Actually, this is the reason why we needed to run ${\mathcal O}(100)$ trajectories to obtain a reasonably stable result. Such large fluctuations have not been seen in the previous simulation of the JIMWLK equation where `already one trajectory gives a good estimate of the final result' \cite{Rummukainen:2003ns}.

 In our opinion, the crucial difference between the two problems which has resulted in such different behaviors of fluctuations is the initial condition. In the jet problem, the initial condition $\langle P\rangle_{\tau=0} =1$ means that there are no partons besides the $q\bar{q}$ pair. In the parlance of saturation physics, the system is initially very `dilute'. The evolution then produces soft gluons which multiply exponentially in $\tau$ according to the BFKL formula \cite{Marchesini:2003nh,Hatta:2008st}. It has been demonstrated that these gluons have very strong number fluctuations \cite{Salam:1995zd} and  spatial correlations \cite{Hatta:2007fg,Avsar:2008ph,Avsar:2009yb}. We thus find it  natural to attribute the observed large values of $\delta P$ to such fluctuations and correlations.
 On the other hand, when solving the JIMWLK equation, one often uses  `dense' or `classically saturated' initial conditions; $\langle S_{\x\y}\rangle_{\tau=0} \ll 1$ for $|\x-\y|$ larger than some value. Since there are many uncorrelated gluons in the system from the beginning, there is little room left for pure--BFKL evolution, i.e., it is suppressed by the saturation effect. Accordingly, fluctuations and correlations can develop only weakly, and  this is consistent with what has been found in the previous simulations. \\

In conclusion, we have demonstrated for the first time the resummation of non-global logarithms at finite $N_c$ to leading logarithmic accuracy. Our study shows that, at least in $e^+e^-$ annihilation and for phenomenologically interesting values of $\tau \lesssim 0.2\sim 0.3$, event--averaged non-global observables may be reliably computed in the mean field  approximation by solving the (modified) BMS   equation  (\ref{finitemean}), or equivalently, by Monte Carlo simulations \cite{Dasgupta:2001sh}. However, the observed large fluctuations imply that the situation may be drastically different for hadron--hadron collisions (cf. \cite{Hatta:2013qj}). Since four partons are involved in hard scattering, one has to deal with multiple products of Wilson lines such as  $\tr (UU^\dagger) \tr (UU^\dagger)$ and $\tr (UU^\dagger UU^\dagger)$ (cf. \cite{Dumitru:2011vk}). Moreover, if there are gluons in the initial and final states, each of them picks up an adjoint Wilson line $\tilde{U}$ which further increases the number of (fundamental) Wilson lines. [Roughly, $\tilde{U}$ acts like the square of $U$.] The proper treatment of fluctuations laid out in this paper is potentially very important for such observables. We leave this problem for future work.

\section*{Acknowledgments}
We thank Tuomas Lappi for useful correspondence. The work of T. U. was supported by the DFG through SFB/TR 9
``Computational Particle Physics''.
The calculations were partially performed on the HP XC3000
at Steinbuch Centre for Computing of KIT.

\end{document}